\documentclass[11pt,a4paper]{article}
\usepackage{amsmath}
\usepackage{amsfonts}
\usepackage{color}
\usepackage{dsfont}
\usepackage{graphicx}
\graphicspath{{figures/}}
\usepackage{dcolumn}
\usepackage{bm}
\usepackage[
bookmarks,bookmarksnumbered,colorlinks=true,anchorcolor=blue,
linkcolor=blue,urlcolor=blue,citecolor=blue,
breaklinks=true]{hyperref}
\usepackage{geometry}
\usepackage{hep}
\usepackage{extarrows}
\usepackage{authblk}
\usepackage{amssymb}
\usepackage[numbers,sort&compress]{natbib}
\usepackage{cite}
\usepackage{pifont}
\usepackage{tikz}
\usepackage{sansmath}
\usepackage{mathrsfs}
\usetikzlibrary{shadings,intersections,snakes}
\usepackage{verbatim}
\numberwithin{equation}{section}   
\usepackage{tikz}  

\date{\today}

\def\s{{\rm s}}

\def\a{{\rm a}}

\def\i{{\rm i}}

\begin{document}

\title{\bf Chiral Anomalous Magnetohydrodynamics in action: effective field theory and holography}

\author[2,3]{Matteo Baggioli \thanks{b.matteo@sjtu.edu.cn}}
\author[1]{Yanyan Bu \thanks{yybu@hit.edu.cn}}
\author[1]{Xiyang Sun \thanks{xysun@stu.hit.edu.cn}}

\affil[1]{\it School of Physics, Harbin Institute of Technology, Harbin 150001, China} 
\affil[2]{\it Wilczek Quantum Center, School of Physics and Astronomy, Shanghai Jiao Tong University, Shanghai 200240, China}
\affil[3]{\it Shanghai Research Center for Quantum Sciences, Shanghai 201315,China}

\maketitle

\begin{abstract}
Chiral Anomalous Magnetohydrodynamics (CAMHD) provides a low-energy effective framework for describing chiral fluids in the presence of dynamical electromagnetic fields and axial anomaly. This theory finds applications across diverse physical systems, including heavy-ion collisions, the early universe, and Weyl/Dirac semimetals. Along with Schwinger-Keldysh (SK) effective theories, holographic models serve as a complementary tool to provide a systematic formulation of CAMHD that goes beyond the weak coupling regime. In this work, we explore holographic models with $U(1)_A \times U(1)$ symmetry, where the electromagnetic $U(1)$ field is rendered dynamical through mixed boundary conditions applied to the bulk gauge field and the axial anomaly is introduced via a Chern-Simons bulk term. Through a detailed holographic SK analysis, we demonstrate that the low-energy effective action derived from this model aligns precisely with the SK field theory proposed by Landry and Liu and, in fact, it generalizes it to scenarios with finite background axial field. This alignment not only validates the holographic model but also paves the way for its use in exploring unresolved aspects of CAMHD, such as the recently proposed chiral magnetic electric
separation wave and nonlinear chiral instabilities.

\end{abstract}

{\let\thefootnote\relax\footnotetext{Authors are ordered alphabetically and should be all considered as co-first authors as well as co-corresponding authors.} }

\newpage


\allowdisplaybreaks

\flushbottom

\section{Introduction}
Hydrodynamics describes late-time and long-distance dynamics of conserved quantities around thermal equilibrium and represents in this sense an effective theory where `fast' degrees of freedom are integrated out \cite{Kovtun:2012rj}. In this limit, a finite temperature chiral fluid in presence of dynamical electromagnetic fields and with Adler-Bell-Jackiw anomaly \cite{Jackiw:2008} is effectively described by chiral anomalous magnethodydrodynamics (CAMHD), that includes the magnetic fields as slow variables together with the other conserved charges. Importantly, the Adler-Bell-Jackiw (ABJ) anomaly breaks the conservation of axial current. Hence, CAMHD is valid only in the limit where axial charge can be treated as an approximately conserved quantity or, in other words, when the axial charge relaxation time is long compared to the typical microscopic timescale -- the \textit{quasi-hydrodynamic} regime \cite{Grozdanov:2018fic}.

In the weak magnetic field regime, CAMHD has been discussed at length in the past \cite{PhysRevD.92.043004,PhysRevD.96.023504,PhysRevD.94.081301,PhysRevD.93.125016,Rogachevskii_2017,PhysRevD.100.065023,sym14091851}. More recently, CAMHD has been considered in the strong field regime and formulated as a full-fledged dissipative effective field theory using the Schwinger-Keldsyh formalism. Das, Iqbal and Poovuttikul \cite{das2022towards} proposed a description in terms of a global one-form symmetry associated to the conservation of the magnetic flux \cite{PhysRevD.95.096003}. In their approach, they were able to reproduce some aspects of chiral anomalous MHD phenomenology from an effective theory viewpoint, including the chiral separation and magnetic effects. Additionally, in that formalism the effects of one-loop level magnetohydrodynamic fluctuations was also studied in \cite{Das:2024efs}. An alternative route has been advanced by Landry and Liu \cite{landry2022systematic} using a more classical formulation in terms of gauge fields. They incorporated the chiral magnetic effect, the chiral separation effect and the chiral electric separation effect in an effective theory language as well. Interestingly, they also predicted the emergence of a novel collective excitation named the \textit{chiral magnetic electric separation wave} \cite{landry2022systematic}.

Despite these two recent developments, a complete effective field theory description of CAMHD, including also velocity and temperature fluctuations, is still missing. Historically, holography has always been a good complementary tool to understand the structure of effective field theories and to build corresponding microscopic models that are valid beyond the weak coupling regime. Chiral hydrodynamics in the presence of strong magnetic fields has been recently discussed in the context of holography and the predictions from hydrodynamics have been matched to the holographic results \cite{ammon2021chiral}. On the other hand, mixed boundary conditions for the $U(1)$ electromagnetic bulk gauge field have been recently employed to make the boundary gauge field dynamical and induce dynamical electromagnetism in the boundary field theory \cite{Domenech:2010nf,Natsuume:2022kic,Ahn:2022azl,baggioli2023sit,Natsuume:2024ril,Jeong:2023las}. This formalism led to the construction of a holographic model of relativistic magnetohydrodynamics, matching the prediction from the hydrodynamic framework \cite{hernandez2017relativistic}, and making the use of bulk higher-form fields not necessary \cite{grozdanov2019generalised}.

In this work, we combine these two recent ideas and build a holographic model of chiral anomalous magnetohydrodynamics. A similar holographic model was recently considered to study the fate of chiral magnetic waves in strongly coupled Weyl semimetals with Coulomb interactions \cite{Ahn:2024ozz}. Our holographic model will be related, up to a dualization process, to the framework recently introduced by Das, Gregory and Iqbal \cite{Das:2022auy} that, on the other hand, uses a higher-form bulk gauge field. The two can be connected by a bulk Hodge-dual transformation, as demonstrated in a simple scenario in \cite{DeWolfe:2020uzb}.

In the probe limit, \textit{i.e.} neglecting the fluctuations of temperature and velocity, the dynamics of the quasi-hydrodynamic degrees of freedom in CAMHD are given by (\textit{e.g.}, \cite{landry2022systematic,das2022towards}):
\begin{align}
    &\partial_{\mu}J^\mu=0 , \label{eq1}\\
    & \partial_\nu F^{\nu\mu}= - J^\mu,\label{eq2}\\
    & \partial_\mu J_5^\mu = 12 \kappa \vec E \cdot \vec B.\label{eq3}
\end{align}
Eq.~\eqref{eq1} is the conservation of the $U(1)$ vector current, Eq.~\eqref{eq2} are the Maxwell equations with the coupling constant set to unity and Eq.~\eqref{eq3} is the conservation of the U(1) axial current broken by the ABJ anomaly with $\kappa$ being the anomaly coefficient. These dynamics will be incorporated into the holographic model using symmetry arguments and the holographic dictionary.

In order to confirm the validity of our holographic model and its nature, we then resort to holographic Schwinger-Keldysh (SK) methods \cite{Glorioso:2018mmw,Crossley:2015tka,haehl2016fluid} that allow us to formally derive  effective action of the boundary field theory. These methods have been already successfully applied in several contexts, \textit{e.g.} \cite{Bu:2020jfo,Bu:2021clf,He:2021jna,Bu:2021jlp,Bu:2022esd,He:2022deg,Ghosh:2022fyo,Jana:2020vyx,Chakrabarty:2019aeu,He:2022jnc,Bu:2022oty,Pantelidou:2022ftm,Baggioli:2023tlc,Bu:2024oyz,Liu:2024tqe}. Spoiling our final results, we anticipate that the boundary effective field theory description of our holographic model for CAMHD is in perfect agreement with the recent proposal by Landry and Liu \cite{landry2022systematic}, of which it provides a generalization in presence of background axial gauge field.

We notice that the introduction of background axial gauge fields could in principle be performed also from a pure EFT perspective. Nevertheless, we emphasize that an important advantage of the holographic framework is the possibility of providing microscopic expressions for the various EFT coefficients. In summary, we view holography as a useful complementary tool that can accompany and help pure EFT descriptions for complex systems.\color{black}\\\\

\textit{Note: Given the large number of symbols, we provide a Table \ref{table1} with all the notations used in the rest of the manuscript in Appendix \ref{note}.}

\section{Effective field theory for chiral anomalous magnethodydrodynamics} \label{basic_idea}

Consider a macroscopic system at finite temperature, whose microscopic degrees of freedom include charged chiral fermions interacting with an electromagnetic field. The microscopic description involves massless QED defined by the Lagrangian
\begin{align}
\mathcal L_{\rm micro} = \bar\psi \i \gamma^\mu (\partial_\mu + \i e a_\mu ) \psi - \frac{1}{4} F_{\mu\nu} F^{\mu\nu}, \label{L_QED}
\end{align}
where $\psi$ is the massless charged fermion field, $a_\mu$ is the electromagnetic gauge potential and $F_{\mu\nu}$ the corresponding field strength. Moreover, $e$ is the fermion charge. The theory described by Eq.~\eqref{L_QED} has a conserved vector current operator and an anomalous chiral current operator
\begin{align}
&\partial_\mu \hat J^\mu =0 \qquad \quad \qquad \qquad \qquad \quad {\rm with}  \qquad \hat J^\mu \equiv \bar\psi \gamma^\mu \psi, \\
&\partial_\mu \hat J_5^\mu = - \frac{e^2}{16\pi^2} \varepsilon^{\mu \nu \alpha \beta} F_{\mu\nu} F_{\alpha\beta} \qquad {\rm with} \qquad \hat J_5^\mu \equiv \bar\psi \gamma^\mu \gamma^5 \psi .\label{continuity_current_operator}
\end{align}
In terms of the microscopic degrees of freedom, the partition function is given by
\begin{align}
Z = \int_{\rho_0} [D \psi] [D a_\mu] e^{i \int_{\mathcal C} d^4x \mathcal L_{\rm micro}}, \label{Z_micro}
\end{align}
where $\rho_0$ is the density matrix of a thermal state. Therefore, in Eq.~\eqref{Z_micro} the integration shall be understood along the Schwinger-Keldysh closed time contour $\mathcal C$.

In the long-wavelength and long-time limits, the effective description of this system is chiral anomalous magnetohydrodynamics (CAMHD). Using the Schwinger-Keldysh (SK) formalism, the partition function in Eq.~\eqref{Z_micro} can be equivalently expressed as
\begin{align}
Z = \int [D a_\mu] [D a_{\rm a \mu}] [D \phi][D\phi_a] [D\varphi][D\varphi_a] e^{i S_{\rm eff}[A_\mu, A_{\rm a \mu}, C_\mu, C_{\rm a \mu}]}. \label{Z_macro}
\end{align}
The effective action $S_{\rm eff}$ depends on the following gauge invariant combinations
\begin{align}
&A_\mu = a_\mu + \partial_\mu \phi, \qquad \qquad  A_{\a\mu}= a_{\a\mu} + \partial_\mu \phi_\a,  \nonumber \\
&C_\mu = \mathcal A_\mu + \partial_\mu \varphi, \qquad \qquad C_{\a\mu} = \mathcal A_{\a\mu} + \partial_\mu \varphi_\a .\label{gauge_invariant_combination}
\end{align}
Here, $\phi$ and $\varphi$ are emergent collective variables which will be used to parameterize the currents' expectation values $J^\mu = \langle \hat J^\mu \rangle$ and $J_5^\mu = \langle \hat J_5^\mu \rangle$. Indeed, $\phi$ and $\varphi$ could be thought of as the dynamical degrees of freedom corresponding to the densities $J^0$ and $J_5^0$, respectively. The latter are usually taken as hydrodynamic variables in formulating chiral MHD. The background axial gauge potential $\mathcal A_\mu$ acts as an external source for the chiral current $J_5^\mu$. All other variables have a subindex $_{\rm a}$ and represent the noisy counterparts characteristic of a finite temperature system. Given Eq.~\eqref{gauge_invariant_combination}, the path integral in Eq.~\eqref{Z_macro} can be further simplified into \cite{landry2022systematic}
\begin{align}
Z = \mathcal N \int [D A_\mu] [D A_{\a \mu}][D\varphi] [D \varphi_\a] e^{\i S_{\rm eff} [A_\mu, A_{\a \mu}, C_\mu, C_{\a \mu}] },
\end{align}
where $\mathcal N$ is an overall constant. Moreover, $A_\mu$ is essentially the macroscopic gauge field in the medium.

In Ref.~\cite{landry2022systematic}, the effective action was split into an anomalous part $I_{\rm anom}$ and an invariant part $I_{\rm inv}$ with respect to axial gauge transformations,
\begin{align}
S_{\rm eff} = I_{\rm anom} + I_{\rm inv},
\end{align}
where
\begin{align}
I_{\rm anom} = c \int \left( \varphi_{\a} F\wedge F + 2 \varphi F \wedge F_\a \right). \label{LL1}
\end{align}
Here, the anomaly coefficient $c = - e^2/(4\pi^2)$. The invariant part $I_{\rm inv} = \int d^4x \mathcal L$ is given by
\begin{align}
\mathcal L = & n A_{\a 0} +n_5 \partial_0 \varphi_\a - H_i B_{\a i} + 2 c \partial_0 \vec A \cdot \vec B \varphi_\a - (2c \mu_5 - b_{00}) B_i A_{\a i} \nonumber \\
& + \i (r^{-1} )_{ij} A_{\a i} \left( T A_{\a j} + \i \partial_0 A_j \right) + \i \kappa_{ij} \partial_i \varphi_\a \left(T \partial_j \varphi_a + \i \partial_j \mu_5 \right) \nonumber \\
& + \i \lambda_{ij} \left((T \partial_i \varphi_\a + \i \partial_i \mu_5) A_{\a j} + \partial_i \varphi_\a (T A_{\a j} + \i \partial_0 A_j) \right) \nonumber \\
& +\gamma_{ij} (\partial_i \varphi_\a \partial_0 A_j - \partial_i \mu_5 A_{\a j}) + \mathcal A_{\a 0} J_5^0 + \mathcal A_{\a i} J_5^i ,\label{LL2}
\end{align}
where
\begin{align}
& n = - \frac{\partial \mathbb F}{\partial A_0}, \qquad n_5 = - \frac{\partial \mathbb F}{\partial \mu_5}, \qquad H_i = \frac{\partial \mathbb F}{\partial B_i} \equiv \frac{B_i}{\mu}, \nonumber \\
& r_{ij} = r_\parallel h_i h_j + r_\perp \Delta_{ij} + r_\times \epsilon_{ijk} h_k, \qquad \Delta_{ij} = \delta_{ij} - h_i h_j ,\nonumber \\
& \kappa_{ij} = \kappa_\parallel h_i h_j + \kappa_\perp \Delta_{ij} + \kappa_\times \epsilon_{ijk} h_k, \qquad 
\lambda_{ij} = \lambda_\parallel h_i h_j + \lambda_\perp \Delta_{ij} + \lambda_\times \epsilon_{ijk} h_k, \nonumber \\
& \gamma_{ij} = \gamma_\parallel h_i h_j + \gamma_\perp \Delta_{ij} + \gamma_\times \epsilon_{ijk} h_k, \qquad 
\lambda_{ij}^{\pm} \equiv \lambda_{ij} \pm \gamma_{ij}, \nonumber \\
& J_5^0 = n_5, \qquad J_5^i = (b_{50} - 2c A_0) B_i -\kappa_{ij} \partial_j \mu_5 - \lambda_{ij}^{-} \partial_0 A_j + \epsilon_{ijk} \partial_j (m B_k).
\end{align}
Here, $\mathbb F$ can be interpreted as the equilibrium free energy density, and $h_i \equiv B_i/B$. All coefficients should be understood as functions of $A_0, \mu_5, B^2$ except for $b_{00}, b_{50}$ that are constant.

Our objective is to reproduce this EFT structure directly from the holographic model. To avoid clutter, all the technical derivations are relegated in the next Section \ref{dersec}. Here, we limit ourselves to report and discuss the main results.

We derive the effective action $S_{\rm eff} = \int d^4x \mathcal L_{\rm eff}$ using the holographic Schwinger-Keldysh formalism \cite{Glorioso:2018mmw}. We organize $\mathcal L_{eff}$ by number of fields
\begin{align}
\mathcal L_{eff} = \mathcal L_{eff}^{(2)} + \mathcal L_{eff}^{(3)} + \mathcal L_{eff}^{(4)}.
\end{align}
In the present work, we focus on the MHD regime assuming that various dynamical variables scale as \cite{landry2022systematic}
\begin{align}
&\partial_0 \sim \mathcal{O}(\varepsilon^2), \quad \partial_i \sim \mathcal{O}(\varepsilon^1), \quad A_0, ~ C_0 \sim \mathcal{O}(\varepsilon^0), \quad A_i,~ C_i \sim \mathcal{O}(\varepsilon^{-1}), \nonumber \\
&A_{\a 0}, ~ C_{\a 0} \sim \mathcal{O}(\varepsilon^2), \quad A_{\a i}, ~ C_{\a i} \sim \mathcal{O}(\varepsilon^1), \label{MHD_regime}
\end{align}
with $\varepsilon$ a `small' parameter controlling the perturbative expansion. The EFT Lagrangian is truncated at order $\mathcal O(\varepsilon^2)$.

The quadratic Lagrangian obtained from the holographic computations reads
\begin{align}
\mathcal L_{\rm eff}^{(2)} & = \frac{2\pi^2}{\beta^2} A_{\a 0} A_0 - \left(1 + \log \frac{\pi T}{L} \right) B_{\a i} B_i + \i \frac{\pi}{\beta^2} A_{\a i} \left( A_{\a i} + \i \beta \partial_0 A_i \right) \nonumber \\
& + \frac{2\pi^2}{\beta^2} C_{\a 0} C_0 - \log \frac{\pi T}{L}\, \mathcal B_{\a i} \mathcal B_i + \i \frac{\pi}{\beta^2} C_{\a i} \left( C_{\a i} + \i \beta \partial_0 C_i \right) \label{L2},
\end{align}
where $\beta=1/T$ with $T$ the system's temperature, and $L$ is the AdS radius. Notice that the coefficients for $B_{\a i} B_i$ and $\mathcal B_{\a i} \mathcal B_i$ are renormalization scheme-dependent \cite{Bu:2020jfo,Glorioso:2018mmw}.

The cubic part of the Lagrangian is given by
\begin{align}
\mathcal L_{\rm eff}^{(3)}  = & 12 \kappa\, \left[ -\varphi_{\a} E_i B_i - \varphi B_{\a i} E_i - \varphi E_{\a i} B_i \right. \nonumber \\
& \left. \qquad - \varphi_{\a} \partial_0 A_i \, B_i + \mu_5 B_i A_{\a i} +  A_0  B_i \mathcal A_{\a i} \right.\nonumber \\
& \left. \qquad  + \mathcal A_i B_i A_{\a 0} +  A_0 \mathcal B_i A_{\a i} + ( \vec E \times \vec{\mathcal A} )_i A_{\a i} \right] \nonumber \\
& + 4 \kappa \,(C_0 C_i \mathcal B_{\a i} +  C_0 C_{\a i} \mathcal B_i + C_{\a 0} C_i \mathcal B_i + \epsilon^{ijk} C_{\a i} C_j \partial_0 C_k) ,\label{L3}
\end{align}
where $\kappa = e^2/(24\pi^2)$ is the anomaly coefficient and $\mu_5 \equiv C_0$ is the axial chemical potential.

We pause to explain our notations. $E_i, B_i$ denote the physical electromagnetic fields and $E_{\a i}, B_{\a i}$ are their noisy counterparts; $\mathcal E_i, \mathcal B_i$ indicate the axial electromagnetic fields and $\mathcal E_{\a i}, \mathcal B_{\a i}$ the associated noisy parts. Notice that we can define the field strengths either via gauge potentials $a_\mu, \mathcal A_\mu, a_{\a \mu}, \mathcal A_{\a \mu}$ or by the gauge invariant combinations $A_\mu, C_\mu, A_{\a\mu}, C_{\a\mu}$. For example, we have $\vec B = \vec \nabla \times \vec a = \vec \nabla \times \vec A$. Likewise, we have $\vec{\mathcal B}=\vec \nabla \times \vec{\mathcal A} $.

We briefly compare our results, Eq.~\eqref{L3}, with the EFT results by Landry and Liu \cite{landry2022systematic}. The first line of Eq.~\eqref{L3} perfectly matches the anomalous part of the EFT action by Landry and Liu \cite{landry2022systematic}, Eq.~\eqref{LL1}. In the second line of Eq.~\eqref{L3}, the first two terms are exactly the $c$-terms in the first line of Eq.~\eqref{LL2}, and the last term corresponds to the $c$-term hidden in $\mathcal A_{\a i} J_5^i$ of Eq.~\eqref{LL2}. The third and fourth lines of Eq.~\eqref{L3} do not appear in the EFT action in \cite{landry2022systematic} as a consequence of the simplifying assumption of vanishing axial field $\mathcal A_\mu =0$. A comparison of the quartic terms in the effective action will be presented below.

The quartic Lagrangian can be further split into a normal part and an anomalous part
\begin{align}
\mathcal L_{\rm eff}^{(4)} = \mathcal L_{\rm normal}^{(4)} + \mathcal L_{\rm anom}^{(4)}.
\end{align}
The normal part is given by
\begin{align}
\mathcal L_{\rm normal}^{(4)} = & (2\alpha_1 + \alpha_2) \left( 16 A_{\a 0} A_0^3  + \frac{32 \rm i}{\pi} A_{\a i} (A_{\a i} + \i \beta \partial_0 A_i) A_0^2 - \frac{8 \rm i }{\pi r_h^2} A_{\a i} (A_{\a i} + \i \beta \partial_0 A_i) B^2 \right) \nonumber \\
& + \alpha_{2}\left[- \frac{16}{r_h^2} \epsilon_{ijk} A_{\a i} \partial_0 A_j B_k A_0 +  \frac{8 \rm i}{\pi r_h^2} A_{\a i} (A_{\a j} + \i \beta \partial_0 A_j ) B_i B_j\right] \nonumber \\
& - \alpha_{1} \frac{32}{3 r_h^2} ( B_{i} B_{ai} A_0^2 + A_{\a 0} A_0 B^2 ) . \label{L4_normal}
\end{align}
The $\alpha$'s are parameters in the holographic action that parameterize the nonlinear corrections to the Maxwell bulk action, see Eq.~\eqref{S0}.

The anomalous part reads
\begin{align}
\mathcal L_{\rm anom}^{(4)} & = \theta_1 (B^2 A_{0} A_{\a 0} + B_i B_{\a i} A_{0} A_{0}  + B^2 C_{0} C_{\a 0} + B_i B_{\a i} C_{0} C_{0} \nonumber \\
& \qquad + \mathcal B^2 A_{0} A_{\a 0} + \mathcal B_i \mathcal B_{\a i} A_{0} A_{0} + \mathcal B^2 C_{0} C_{\a 0} + \mathcal B_i \mathcal B_{\a i} C_{0} C_{0}  ) \nonumber \\
& + 2 \theta_1 (B_i \mathcal B_{\a i} A_{0} C_{0} + B_{\a i} \mathcal B_i A_{0} C_{0} + B_i \mathcal B_i A_{\a 0} C_{0} +  B_i \mathcal B_i A_{0} C_{\a0}) \nonumber \\
& - \theta_2 \epsilon^{ijk} (\partial_0 A_{j} B_k A_{0} A_{\a i} + \partial_0 A_{j} B_k C_{0} C_{\a i} + \partial_0 C_{j} B_k C_{0} A_{\a i} + \partial_0 C_{j} B_k A_{0} C_{\a i} \nonumber \\
&\qquad \quad \,\,  + \partial_0 C_{j} \mathcal B_k A_{0} A_{\a i} + \partial_0 A_{j} \mathcal B_k C_{0} A_{\a i} + \partial_0 A_{j} \mathcal B_k A_{0} C_{\a i} +  \partial_0 C_{j} \mathcal B_k C_{0} C_{\a i} ) \nonumber \\
& + \i \theta_3 ( B^2\delta_{ij} + B_i B_j + \mathcal B^2\delta_{ij} + \mathcal B_i \mathcal B_j )\left[ \left( A_{\a i} + \i \beta \partial_0 A_{i} \right) A_{\a j} + \left( C_{\a i} + \i \beta \partial_0 C_{i} \right) C_{\a j} \right] \nonumber \\
& + \i \theta_3 ( 2 B_k \mathcal B_k \delta_{ij} + B_i \mathcal B_j  + \mathcal B_i B_j   )\left[ \left( A_{\a i} + \i \beta \partial_0 A_{i}\right) C_{\a j} + \left( C_{\a i} + \i \beta \partial_0 C_{i} \right) A_{\a j} \right], \label{L4_anom1}
\end{align}
where
\begin{align}
\theta_1 = \frac{9 ( -1+2 \log2)}{2 r_h^2}\kappa^2 , \quad
\theta_2 = \frac{9 \log2}{ r_h^2}\kappa^2, \quad
\theta_3 = \frac{9 \log2}{4 \pi  r_h^2} \kappa^2. \label{theta_values}
\end{align}
The EFT action $S_{\rm eff}$ shall satisfy dynamical KMS symmetry \cite{Glorioso:2017fpd,Glorioso:2018wxw}:
\begin{align}
S_{\rm eff}[A_\mu, A_{\a \mu}, C_\mu, C_{\a \mu}] = S_{\rm eff}[\tilde A_\mu, \tilde A_{\a \mu}, \tilde C_\mu, \tilde C_{\a \mu}], \label{KMS1}
\end{align}
where
\begin{align}
& \tilde A_\mu(-x) = A_\mu(x), \quad \qquad \tilde A_{\a \mu}(-x) = A_{\a \mu}(x) + \i \beta \partial_0 A_\mu(x), \nonumber \\
& \tilde C_\mu(-x) = -C_\mu(x), \qquad \,\, \tilde C_{\a \mu}(-x) = - \left[C_{\a \mu}(x) + \i \beta \partial_0 C_\mu(x) \right]. \label{KMS2}
\end{align}
We would like to stress that our results Eqs.~\eqref{L2}-\eqref{L3}-\eqref{L4_normal}-\eqref{L4_anom1} are presented in a transparent KMS-invariant fashion.

We now compare our results with equations (4.4) and (4.8) in Ref.~\cite{landry2022systematic}, which are quoted in Eq.~\eqref{LL1} and Eq.~\eqref{LL2}. To this end, we switch off background axial field by setting $\mathcal A_\mu =0$, and keep terms linear in $\mathcal A_{\a \mu}$ only. This will simplify the action in many ways: (I) in Eq.~\eqref{L2} the term $\mathcal B_{\a i} \mathcal B_i$ is dropped; (II) only the first two lines in equation \eqref{L3} survive; (III) in \eqref{L4_anom1}, the second line, the third line (except the first term therein), the fifth line and the last line are dropped. Finally, in Eq.~\eqref{L4_anom1} we shall replace $C_{\a \mu} = \mathcal A_{\a \mu} + \partial_\mu \varphi_a$. Eventually, we can obtain the correspondence between the EFT coefficients presented in \cite{landry2022systematic} and our holographic data as follows:
\begin{equation}
\begin{cases}
& \frac{2 \pi^2}{\beta^2}A_0 + 32(\alpha_{1} + \frac{1}{2} \alpha_2)A_0^3 - \frac{32 \beta^2}{3 \pi^2}\alpha_{1} A_0 B^2 + \theta_1 B^2 A_{0}\rightarrow n ; \\
& 1 + \log \frac{\pi T}{L} + \frac{32 \beta^2}{3 \pi^2}\alpha_{1}A_0^2 - \theta_1(A_{0}^2 + \mu_5^2) \rightarrow \frac{1}{\mu};  \\
& \frac{2 \pi^2}{\beta^2}C_0 + \theta_1(B^2 \mu_5)\rightarrow n_5 ; \\
& 2 \theta_1 \epsilon^{ijk} \partial_{j}(A_{0} \mu_5 B_{k}) \rightarrow \epsilon^{ijk} \partial_{j} (m B_{k});  \\
& \frac{16 \beta^2}{\pi^2} \alpha_{2} \epsilon^{ijk}  A_0 B_k + \epsilon^{ijk} \theta_2 ( B_{k} A_{0} )\rightarrow \epsilon^{ijk} (r^{-1})_{\times} h_{k} ; \\
& \theta_2 \epsilon^{ijk} (B_{k} \mu_5) \rightarrow  \epsilon^{ijk} (\lambda_{\times}-\gamma_{\times}) h_{k} ; \\
& - \theta_2 \epsilon^{ijk}  (B_{k} \mu_5) \rightarrow  \epsilon^{ijk} (\lambda_{\times}+\gamma_{\times}) h_{k} ; \\
& \theta_2 \epsilon^{ijk}  (B_{k} A_{0}) \rightarrow \epsilon^{ijk} \kappa_{\times} h_{k} ; \\
& \frac{ \pi}{\beta} \delta_{ij} +  \frac{16 \beta }{\pi}(\alpha_{1} + \frac{1}{2} \alpha_2)(4 A_0^2 - \frac{ \beta^2 B^2}{ \pi^2}) \delta_{ij} + \theta_3 \beta B^2 \delta_{ij}  \rightarrow  (r^{-1})_{\perp} \delta_{ij};  \\
& \frac{8 \beta^3}{\pi^3} \alpha_{2} B_{i} B_{j} +  \theta_3 \beta B_{i} B_{j}  \rightarrow \left[(r^{-1})_{\parallel} - (r^{-1})_{\perp} \right] h_{i} h_{j} ; \\
& \frac{ \pi}{\beta} \delta_{ij} + \theta_3 \beta B^2 \delta_{ij} \rightarrow \kappa_{\perp} \delta_{ij} ; \\
& \theta_3 \beta B_{i} B_{j} \rightarrow (\kappa_{\parallel} - \kappa_{\perp}) h_{i} h_{j}.
\end{cases}
\end{equation}
Here, we remind that $h_i \equiv B_i/B$. We also recall that $A_0$ is the vector chemical potential while $C_0$ is the axial chemical potential and that $\theta_1,\theta_2,\theta_3$ all vanish in the limit $\kappa \rightarrow 0$, \textit{i.e.} by switching off the chiral anomaly. Also, $\alpha_{1,2}$ are the parameters in the holographic model that control the nonlinear corrections to the Maxwell bulk action, see Eq.~\eqref{S0}.

The SK effective action $S_{\rm eff}$ allows us also to directly obtain the expectation values of the current operators defined in Eq.~\eqref{continuity_current_operator},
\begin{align}
\mathcal J^\mu \equiv \frac{\delta S_{\rm eff}}{\delta a_{\a \mu}}, \qquad \qquad J_5^\mu \equiv  \frac{\delta S_{\rm eff}}{\delta \mathcal A_{\a \mu}}, \label{currents_definition_EFT}
\end{align}
where $\mathcal J^\mu = \partial_\nu F^{\nu\mu} + J^\mu$. Here, the presence of $\partial_\mu F^{\mu\nu}$ is simply due to the added kinetic term for the dynamical electromagnetic field $a_\mu$. We notice that the currents defined above are the \textit{consistent currents} that can be made fully covariant by adding the Chern-Simons current (see \textit{e.g.} \cite{Landsteiner:2016led}). Moreover, within the EFT framework, the dynamics of the system are derived from the variation of $S_{\rm eff}$ with respect to dynamical fields $A_{\a\mu}$ and $\varphi_\a$, leading to the expected expressions:
\begin{align}
& \frac{\delta S_{\rm eff}}{\delta A_{\a\mu}} = 0 \,\Longrightarrow \mathcal J^\mu = 0\,\,\Longleftrightarrow \,\,\partial_\nu F^{\nu\mu} = - J^\mu, \nonumber \\
& \frac{\delta S_{eff}}{\delta \varphi_\a} = 0 \,\Longrightarrow\,\, \partial_\mu J^\mu_5 = - \frac{e^2}{16\pi^2} \varepsilon^{\mu\nu\alpha\beta} F_{\mu\nu} F_{\alpha\beta}. \label{variational_EFT}
\end{align}
In the context of MHD, the first equation in \eqref{variational_EFT} is considered as the constitutive relation for the electric field, which could be recast into an evolution equation for the magnetic field; on the other hand, the second equation in \eqref{variational_EFT} is simply the non-conservation law for the chiral current. In fact, owing to the truncation to order $\mathcal O(\varepsilon^2)$ characterizing the MHD regime \eqref{MHD_regime}, the Maxwell equations in \eqref{variational_EFT} simplify to
\begin{align}
J^0 =0, \qquad (\vec \nabla \times \vec B)^i = J^i.
\end{align}

From \eqref{currents_definition_EFT}, the explicit expressions for the vector and axial currents are
\begin{align}
	J^0 =& \frac{2 \pi^2}{\beta^2} A_0 + 12 \kappa \mathcal A_i B_i + (2\alpha_{1} +  \alpha_{2}) 16 A_0^3 - \frac{32 \beta^2}{3 \pi^2} \alpha_1 A_0 B^2 \nonumber \\
    & + \theta_1 (A_0 B^2 + \mathcal B^2 A_0 + 2 B_i \mathcal B_i C_0), \nonumber \\
	J^i =&  -\log{\frac{\pi T}{L}} (\vec \nabla \times \vec B)_i - \frac{\pi}{\beta}\partial_0 A_{i} + 12 \kappa \left[ \mu_5 B_i + A_0 \mathcal B_i + ( \vec E \times \vec C )_i \right] \nonumber \\
	& - (2 \alpha_{1} +  \alpha_{2}) \frac{8 \beta}{\pi} \left( 4 A_0^2 - \frac{ \beta^2}{\pi^2} B^2 \right) \partial_{0} A_i - \alpha_1 \frac{32 \beta^2}{3 \pi^2} \epsilon_{ijk} \partial_{j} (A_0^2 B_k) \nonumber \\
	&  - \alpha_{2} \frac{16 \beta^2}{\pi^2} \epsilon_{ijk} \partial_{0} A_j B_k  A_0 - \alpha_2 \frac{8 \beta^3}{ \pi^3}B_i B_j \partial_{0} A_j + \theta_1 \epsilon_{ijk} \partial_{j}\left[ B_k(A_0 + C_0)^2 \right] \nonumber \\
	& - \theta_2 \epsilon_{ijk}( A_0 \partial_{0} A_j B_k + C_0 \partial_{0} C_j B_k + A_0 \partial_{0} C_j \mathcal B_k + C_0 \partial_{0} A_j \mathcal B_k)  \nonumber \\
	& - \theta_3 \beta (B^2 \delta_{ij} + B_i B_j + \mathcal B^2 \delta_{ij} + \mathcal B_i \mathcal B_j) \partial_{0} A_j \nonumber \\
	& - \theta_3 \beta (2 B_k \mathcal B_k \delta_{ij} + B_i \mathcal B_j + \mathcal B_i B_j) \partial_{0} C_j, \nonumber \\
	J_5^0 = & \frac{2 \pi^2}{\beta^2} C_0 + 4 \kappa  C_i \mathcal B_i  + \theta_1 (C_0 B^2 + \mathcal B^2 C_0 + 2 B_i \mathcal B_i A_0), \nonumber \\
	J_5^i = &  -\log{\frac{\pi T}{L}} (\vec \nabla \times \vec{\mathcal B})_i - \frac{ \pi}{\beta} \partial_0 C_{i} + 4 \kappa [2 C_0 \mathcal B_i + 3 A_0  B_i + (\vec{\mathcal E} \times \vec C)_i]  \nonumber \\
	& + \theta_1 \epsilon_{ijk} \partial_{j}[\mathcal B_k(A_0+ C_0)^2 ] - \theta_2 \epsilon_{ijk}( C_0 \partial_{0} A_j B_k + A_0 \partial_{0} C_j B_k + A_0 \partial_{0} A_j \mathcal B_k \nonumber \\
	& + C_0 \partial_{0} C_j \mathcal B_k)  - \theta_3 \beta (B^2 \delta_{ij} + B_i B_j + \mathcal B^2 \delta_{ij} + \mathcal B_i \mathcal B_j) \partial_{0} C_j\nonumber \\
	& - \theta_3 \beta (2 B_k \mathcal B_k \delta_{ij} + B_i \mathcal B_j + \mathcal B_i B_j) \partial_{0} A_j,
\end{align}
where we have dropped the noise contributions. We can make the further replacement $\partial_0 A_i = \partial_i A_0 - E_i$ and $\partial_0 C_i = \partial_i \mu_5 - \mathcal E_i$ so that relevant terms in the currents above become more transparent. In this direction, we recall that $A_0$ is the vector chemical potential while $C_0$ is the axial chemical potential. Finally, we again remind the readers that the currents above are the consistent currents, differing from the covariant ones by the Chern-Simons current \cite{Landsteiner:2016led,Amoretti:2022vxq}.

\section{Holographic derivation}\label{dersec}

Here, we provide all the technical details of the holographic computations and derive the effective action presented in the previous Section.

\subsection{Holographic setup}

The holographic model involves a vector gauge field $V_M$ and an axial gauge field $L_M$ whose dynamics are described by the following bulk action
\begin{align}
S_0 = \int d^5x \sqrt{-g} & \left[ - \frac{1}{4} \mathds{V}_{MN} \mathds{V}^{MN}  - \frac{1}{4} \mathds{L}_{MN} \mathds{L}^{MN} + \frac{\kappa}{2} \epsilon^{MNPQR} L_M \left( 3 \mathds{V}_{NP} \mathds{V}_{QR} + \mathds{L}_{NP} \mathds{L}_{QR} \right) \right. \nonumber \\
&\,\, \left. + \alpha_1 \left( \mathds{V}_{MN} \mathds{V}^{MN} \right)^2 + \alpha_2 \mathds{V}_{AB} \mathds{V}^{BC} \mathds{V}_{CD}\mathds{V}^{DA} \right], \label{S0}
\end{align}
where $\mathds{V}_{MN} = \nabla_M V_N - \nabla_N V_M$, $\mathds{L}_{MN} = \nabla_M L_N - \nabla_N L_M$, and $\epsilon^{MNPQR}$ represents the Levi-Civita tensor in 5D AdS space. The vector field and axial field interact via a gauge Chern-Simons term inducing a chiral anomaly in the boundary theory. In addition, quartic terms for the vector gauge field $V_M$ are included to generate non-anomalous nonlinear terms in the boundary action, see \eqref{L4_normal}. Practically, the anomaly coefficient $\kappa$ and the coupling constants $\alpha_{1,2}$ are assumed to be much smaller than one, validating our perturbative calculations. One may wonder whether we need to add further higher-order terms in the bulk model \eqref{S0} in order to generate further non-linearities in the boundary action. Indeed, this is unnecessary as seen from \eqref{L4_anom1}: quartic terms in the boundary action can be generated by cubic interactions in the bulk (i.e., chiral anomaly terms $\sim \kappa$). More generally, an interacting term in the bulk will generate same order non-linearities in the boundary action (via contact-type Witten diagram in the AdS) or induce higher-order non-linearities in the boundary action (via exchange-type Witten diagram in the AdS).

\begin{figure} [htbp!]
		\centering
		\begin{tikzpicture}[]
		\draw[cyan, ultra thick] (-5.5,0)--(-2.8,0);
		\draw[cyan, ultra thick] (-1.2,0)--(1.5,0);
		\draw[cyan, ultra thick] (-2.81,0.019) arc (-180:0:0.8);
		\draw[cyan, ->,very thick] (-1,0)--(0.2,0);
		\draw[cyan, ->,very thick] (-4.2,0)--(-4,0);
		\draw[fill] (-5.5,0) circle [radius=0.05];
		\node[above] at (-5.5,0) {\small $\infty_2$};
		\draw[fill] (1.5,0) circle [radius=0.05];
		\node[above] at (1.5,0) {\small $\infty_1$};
		\draw[fill,red] (-2,0) circle [radius=0.05];
		\node[above] at (-2,0) {\small $r_h$};				
		
		\draw[-to,thick] (1.85,0)--(2.45,0);

		\node[below] at (9.5,-0.1) {\small Re$(r)$};
		\draw[->,ultra thick] (2.9,0)--(9.6,0);
		\node[right] at (3.8,1) {\small Im$(r)$};
		\draw[->,ultra thick] (3.6,-1)--(3.6,1.2);
		\draw[cyan, very thick] (5.2,-0.18) arc (165:-165:-0.7);
		\draw[cyan, very thick] (5.2,0.2)--(8.3,0.2);
		\draw[cyan, very thick] (5.2,-0.2)--(8.3,-0.2);
		\draw[cyan, ->,very thick] (6,-0.2)--(7,-0.2);
		\draw[cyan, <-,very thick] (6,0.2)--(7,0.2);
		\draw[fill] (3.6,0) circle [radius=0.05];
		\draw[fill] (8.3,0.2) circle [radius=0.05];
		\node[below] at (8.3,-0.15) {\small $\infty_1 $};
		\draw[fill] (8.3,-0.2) circle [radius=0.05];
		\node[above] at (8.3,0.15) {\small $\infty_2 $};
		\draw[fill , red ] (4.5,0) circle [radius=0.07];
		\node[below] at (4.5,0) {\footnotesize $r=r_h $};
		\draw[cyan, thick,<->] (4.52,0.08)--(4.8,0.62);
		\node[above] at (4.5, 0.2) {\footnotesize $\epsilon$};
		\end{tikzpicture}
		\caption{From complexified double AdS (left) \cite{Crossley:2015tka} to the holographic SK contour (right) \cite{Glorioso:2018mmw}. The two horizontal legs overlap with the real axis.} \label{rcontour}
\end{figure}
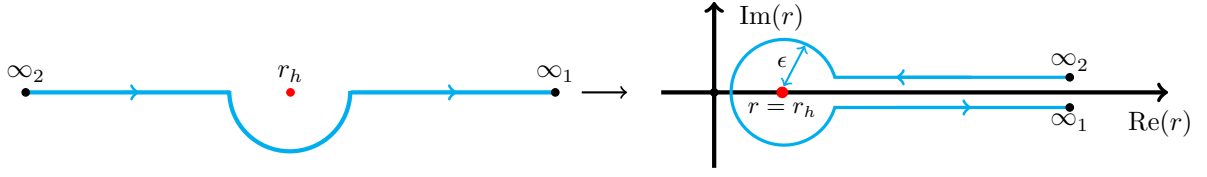

We consider a Schwarzschild-AdS$_5$ black brane geometry
\begin{align}
ds^2 = g_{MN} dx^M dx^N = 2dv dr - r^2f(r) dv^2 + r^2 \delta_{ij} dx^i dx^j,
\end{align}
where $f(r) = 1-r_h^4/r^4$, $x^M = (r, x^\mu)$ with $r$ the holographic radial coordinate, and $x^\mu = (v, x^i)$ represent the boundary spacetime coordinates. The location of the event horizon is at $r=r_h$ which is related to the boundary temperature via $r_h = \pi T$. In order to derive the SK effective action on the boundary, we use the holographic prescription \cite{Glorioso:2018mmw} for the SK closed time contour: analytically continue the radial coordinate $r$ around the horizon and then double it, see Figure \ref{rcontour}.

The bulk equations of motion are
\begin{align}
EV^M = 0, \qquad \qquad \qquad EL^M =0 , \label{bulk_eom}
\end{align}
where
\begin{align}
& EV^M \equiv \nabla_N \mathds{V}^{NM}  +  3 \kappa  \epsilon^{MNPQR} \mathds{V}_{NP} \mathds{L}_{QR} + 8 \nabla_N \left( \alpha_1 \mathds{V}^{NM}  \mathds{V}_{AB} \mathds{V}^{AB} + \alpha_2 \mathds{V}^{NA} \mathds{V}_{AB} \mathds{V}^{BM} \right) ,\nonumber \\
& EL^M \equiv \nabla_N \mathds{L}^{NM} + \frac{3 \kappa }{2} \epsilon^{MNPQR} \left( \mathds{V}_{NP} \mathds{V}_{QR} + \mathds{L}_{NP} \mathds{L}_{QR} \right).
\end{align}
We will take the following radial gauge conditions \cite{Bu:2020jfo,Bu:2021clf}
\begin{align}
V_r = - \frac{V_0}{r^2 f(r)}, \qquad \qquad  L_r = - \frac{L_0}{r^2 f(r)}.
\end{align}
Near the AdS boundary $r=\infty_\s$ with $\s=1~{\rm or}~2$, we have
\begin{align}
& V_\mu(r \to \infty_\s, x^\alpha) = A_{\s\mu}(x^\alpha) + \cdots + \frac{\mathfrak J_{\s\mu}(x^\alpha)}{r^2} + \cdots, \nonumber \\
& L_\mu(r \to \infty_\s, x^\alpha) = C_{\s\mu}(x^\alpha) + \cdots + \frac{\mathfrak L_{\s\mu}(x^\alpha)}{r^2} + \cdots  \label{VL_bdy_expansion}
\end{align}
In the (r,a)-basis, the boundary data $A_{\s\mu}$ and $C_{\s\mu}$ will be combined into \eqref{gauge_invariant_combination}
\begin{align}
& A_\mu \equiv \frac{1}{2} \left( A_{1\mu} + A_{2\mu} \right), \qquad \qquad  A_{\a\mu} \equiv A_{1\mu} - A_{2\mu}, \nonumber \\
& C_\mu \equiv \frac{1}{2} \left( C_{1\mu} + C_{2\mu} \right),  \qquad \qquad C_{\a\mu} \equiv C_{1\mu} - C_{2\mu}.
\end{align}
We will work in the saddle point approximation for the bulk theory. Therefore, the derivation of the boundary effective action involves only the solutions of the bulk equations of motion \eqref{bulk_eom}. Yet, we shall take a partially on-shell prescription \cite{Bu:2014ena,Crossley:2015tka,Bu:2020jfo}: we will solve the dynamical components of the bulk equations \eqref{bulk_eom}
\begin{align}
& EV^0 - \frac{EV^r}{r^2f(r)} =0, \qquad \qquad EV^i =0, \nonumber \\
& EL^0 - \frac{EL^r}{r^2f(r)} =0, \qquad \qquad EL^i =0, \label{dynamical_eom}
\end{align}
while leave aside the constraint components corresponding to $M=r$ in \eqref{bulk_eom}.

Our boundary conditions are as follows. For the axial bulk field $L_\mu$, we will impose Dirichlet condition, \textit{i.e.}, fix $C_{\s\mu}$ in \eqref{VL_bdy_expansion}. In contrast, for the vector bulk field $V_\mu$, we will take an alternative quantization scheme by fixing the normalizable mode $\mathfrak J_{\s\mu}$ in \eqref{VL_bdy_expansion}. In this way, we will  have a ``dynamical'' gauge field $a_{\s\mu}$ and a ``background'' axial gauge field $\mathcal A_{\s\mu}$ on the boundary. Thus, after solving \eqref{dynamical_eom}, we shall obtain the following functional relations
\begin{align}
A_{\s\mu} = A_{\s\mu}[\mathfrak J_{\s\mu}], \qquad \qquad  \mathfrak L_{\s\mu} = \mathfrak L_{\s\mu}[C_{\s\mu}].
\end{align}

Indeed, the bulk action \eqref{S0} shall be supplemented with the following boundary term
\begin{align}
S_{\rm bdy} = \int d^4x \sqrt{-\gamma} \left[ -\frac{\log r}{4} \mathds{V}_{\mu\nu} \mathds{V}^{\mu\nu} -\frac{\log r}{4} \mathds{L}_{\mu\nu} \mathds{L}^{\mu\nu} -\frac{1}{4} \mathds{V}_{\mu\nu} \mathds{V}^{\mu\nu} \right]\Bigg|_2^1 ,\label{S_bdy}
\end{align}
where the first two parts are counter-term action for removing divergences at AdS boundary based on a minimal subtraction scheme, and the last part represents a kinetic term for dynamical gauge field on the boundary. This last term could also be reabsorbed in the boundary conditions for the bulk gauge field that, in that way, become mixed boundary conditions rather than alternative \cite{baggioli2023sit}. The inclusion of this last term guarantees that the bulk variational problem (in the on-shell sense) is well-defined
\begin{align}
\delta (S_0 + S_{\rm bdy}) = \int d^4x \left( A_{\s \mu} \delta \mathfrak J_\s^\mu + \mathfrak L_\s^\mu \delta C_{\s\mu} \right).
\end{align}
It is important to note that, because of the structure of the EFT, we can recycle previous results for the vector bulk field $V_\mu$ well studied for the case of charge diffusion, in which a Dirichlet condition was imposed for $V_\mu$. This can be simply understood as follows. Inverting $A_{\s\mu} = A_{\s\mu}[\mathfrak J_{\s\mu}]$ gives $\mathfrak J_{\s\mu} = \mathfrak J_{\s\mu}[A_{\s\mu}]$. Then, a boundary action of the form $\int d^4x A_{\s \mu} \mathfrak J_{\s}^\mu $ can be viewed as a functional of a dynamical gauge field $A_{\s\mu}$, which is the same as that of the case of charge diffusion.

Additionally, in order to fully determine the bulk solutions, we have to impose an additional condition for the time-components of bulk gauge fields \cite{Glorioso:2018mmw}
\begin{align}
V_0(r= r_h - \epsilon, x^\alpha) =0, \qquad \qquad L_0(r=r_h-\epsilon, x^\alpha) =0 ,\label{horizon_condition}
\end{align}
which corresponds to the chemical shift symmetry for boundary theory \cite{Glorioso:2018mmw,Crossley:2015evo,Knysh:2024asf}.

Once the dynamical bulk equations \eqref{dynamical_eom} are solved, we compute the partially on-shell bulk action by plugging the bulk solutions into \eqref{S0} and performing the radial integration, yielding
\begin{align}
S_{\rm eff} = S_0 |_{\rm p.o.s} + S_{\rm bdy}, \label{pos_action}
\end{align}
where $S_0 |_{\rm p.o.s}$ stands for the partially on-shell bulk action.

\subsection{Bulk solutions}
In this Section we solve the dynamical equations \eqref{dynamical_eom} and then compute the partially on-shell bulk action \eqref{pos_action}.

\subsubsection{The scheme of perturbative expansion}

Since Eqs.~\eqref{dynamical_eom} are highly nonlinear, we search for perturbative solutions through a double expansion scheme.

First, recall that the anomaly coefficient $\kappa$, and coupling constants $\alpha_{1,2}$ are assumed to be much smaller than unity. Thus, we search for perturbative solutions to the bulk equations \eqref{dynamical_eom} of the form
\begin{align}
V_\mu = V_\mu^{(0)} + \lambda\, V_\mu^{(1)} + \cdots, \qquad \qquad  L_\mu = L_\mu^{(0)} + \lambda\, L_\mu^{(1)} + \cdots \label{lambda-expansion}
\end{align}
where we introduced a bookkeeping parameter $\lambda$ to assist the perturbative expansion. Moreover, we assume the scaling $\lambda \sim \kappa \sim \sqrt{\alpha_{1,2}}$.This assumption is mainly motivated by the quartic terms in the boundary action. From the results \eqref{L4_normal}-\eqref{theta_values}, we see that there are two type of contributions to quartic terms in the boundary action: those generated by the chiral anomaly effect having an overall coefficient $\kappa^2$, and those generated by $\alpha$-terms in bulk action \eqref{S0} having an overall coefficient $\alpha_{1,2}$. Thus, from the perspective of quartic terms in the boundary action, we made the scaling assumption $\kappa \sim \sqrt{\alpha_{1,2}}$.
The leading order solutions $V_\mu^{(0)}$ and $L_\mu^{(0)}$ obey homogeneous partial differential equations (PDEs), which are the dynamical components of the free Maxwell equations in Schwarzschild-AdS$_5$. The nonlinear solutions $V_\mu^{(1)}$ and $L_\mu^{(1)}$ obey similar PDEs as those of $V_\mu^{(0)}$ and $L_\mu^{(0)}$, with nontrivial sources to be built from leading order solutions $V_\mu^{(0)}$ and $L_\mu^{(0)}$.

Second, at each order in the expansion \eqref{lambda-expansion} we consider a boundary derivative expansion
\begin{align}
& V_\mu^{(m)} = V_\mu^{(m)(0)} + \gamma\, V_\mu^{(m)(1)} + \gamma^2 \, V_\mu^{(m)(2)} + \cdots, \nonumber \\
& L_\mu^{(m)} = L_\mu^{(m)(0)} + \gamma\, L_\mu^{(m)(1)} + \gamma^2 \, L_\mu^{(m)(2)} + \cdots \label{derivative_expansion}
\end{align}
where $\gamma \sim \partial_\mu$ is assumed to assist the derivative expansion. Working in the MHD regime \eqref{MHD_regime}, the derivative expansion \eqref{derivative_expansion} could be simplified.

Through the double expansion \eqref{lambda-expansion} and \eqref{derivative_expansion}, the nonlinear PDEs of \eqref{dynamical_eom} turn into a set of ordinary differential equations which we schematically write here
\begin{align}
& \partial_r\left[r^3 \partial_r V_0^{(m)(n)} \right] = j_0^{(m)(n)}, \qquad \qquad \partial_r\left[r^3 f(r)\partial_r V_i^{(m)(n)} \right] = j_i^{(m)(n)}, \nonumber \\
& \partial_r\left[r^3 \partial_r L_0^{(m)(n)} \right] = l_0^{(m)(n)}, \qquad \qquad \,\,\, \partial_r\left[r^3 f(r)\partial_r L_i^{(m)(n)} \right] = l_i^{(m)(n)} .\label{dynamical_eom_perturb}
\end{align}
Here, the source terms can be read off by plugging \eqref{derivative_expansion} into \eqref{dynamical_eom}, and they are indeed functionals of the lower order solutions.

\subsubsection{The perturbative solutions} \label{perturb_solutions}

For the leading order terms $V_\mu^{(0)}, L_\mu^{{0}}$ in \eqref{lambda-expansion}, it is possible to recycle the analytical solutions that we obtained previously \cite{Baggioli:2023tlc,Bu:2024oyz}.

At the lowest order $\mathcal O(\lambda^0 \gamma^0)$, we have
\begin{align}
& V_0^{(0)(0)}(r) = A_{\s 0} \left(1- \frac{r_h^2}{r^2} \right), \qquad r\in[r_h-\epsilon, \infty_\s), \nonumber \\
& L_0^{(0)(0)}(r) = C_{\s 0} \left(1- \frac{r_h^2}{r^2} \right), \qquad r\in[r_h-\epsilon, \infty_\s), \nonumber \\
& V_i^{(0)(0)}(r) = A_{2i} + \frac{A_{\a i}}{2\i \pi} \log \frac{r^2-r_h^2}{r^2 +r_h^2}, \nonumber \\
& L_i^{(0)(0)}(r) = C_{2i} + \frac{C_{\a i}}{2\i \pi} \log \frac{r^2-r_h^2}{r^2 +r_h^2}. \label{Vmu_Lmu_00}
\end{align}

At the next order $\mathcal O(\lambda^0 \gamma^1)$, we have
\begin{align}
V_0^{(0)(1)}(r) &= \frac{\partial_0 A_{\s 0}}{4r_h} \left( 1- \frac{r_h^2}{r^2} \right) \left[ \pi - 2 \arctan\left( \frac{r}{r_h}\right) + \log \frac{r+r_h}{r-r_h} \right], \quad r\in[r_h-\epsilon, \infty_\s), \nonumber \\
L_0^{(0)(1)}(r) &= \frac{\partial_0 C_{\s 0}}{4r_h} \left( 1- \frac{r_h^2}{r^2} \right) \left[ \pi - 2 \arctan\left( \frac{r}{r_h}\right) + \log \frac{r+r_h}{r-r_h} \right], \quad r\in[r_h-\epsilon, \infty_\s), \nonumber \\
V_i^{(0)(1)}(r) &= \frac{\partial_0 A_{2i}}{4r_h} \left[ \pi - 2 \arctan\left( \frac{r}{r_h} \right) + 2\log(r+r_h) - \log (r^2 +r_h^2) \right] \nonumber \\
& - \frac{\partial_0 A_{\a i}}{8\pi r_h} \left[ -(2 - \i) \pi - 2 \i \arctan\left( \frac{r}{r_h} \right) - \i \log \frac{r-r_h}{r+r_h}  \right] \log \frac{r^2 -r_h^2}{r^2 + r_h^2}, \nonumber \\
L_i^{(0)(1)}(r) &= \frac{\partial_0 C_{2i}}{4r_h} \left[ \pi - 2 \arctan\left( \frac{r}{r_h} \right) + 2\log(r+r_h) - \log (r^2 +r_h^2) \right] \nonumber \\
& - \frac{\partial_0 C_{\a i}}{8\pi r_h} \left[ -(2 - \i) \pi - 2 \i \arctan\left( \frac{r}{r_h} \right) - \i \log \frac{r-r_h}{r+r_h}  \right] \log \frac{r^2 -r_h^2}{r^2 + r_h^2}. \label{Vmu_Lmu_01}
\end{align}

While $V_\mu^{(0)(2)}, L_\mu^{(0)(2)}$ were previously obtained analytically \cite{Bu:2020jfo}, they are too lengthy to be recorded here. Nevertheless, we can present them in a compact form without working out the radial integral. In fact, this treatment can be straightforwardly applied to nonlinear corrections such as $V_\mu^{(1)}$ and $L_\mu^{(1)}$.

For the time-components of the bulk gauge fields, the higher order solutions can be formally written as \cite{Bu:2021clf,Baggioli:2023tlc}
\begin{align}
& V_0^{(m)(n)}(r) = \int_{\infty_\s}^r \left[ \frac{1}{x^3} \int_{\infty_\s}^x j_0^{(m)(n)}(y) dy + \frac{c_\s^{(m)(n)}}{x^3} \right] dx, \quad r\in[r_h-\epsilon, \infty_\s), \nonumber \\
& L_0^{(m)(n)}(r) = \int_{\infty_\s}^r \left[ \frac{1}{x^3} \int_{\infty_\s}^x l_0^{(m)(n)}(y) dy + \frac{d_\s^{(m)(n)}}{x^3} \right] dx, \quad \,\, r\in[r_h-\epsilon, \infty_\s), \label{V0_L0_mn}
\end{align}
where the integration constants $c_\s^{(m)(n)}$ and $d_\s^{(m)(n)}$ are determined by the vanishing horizon condition \eqref{horizon_condition}.

For spatial components of the bulk fields, the higher order solutions can be constructed via the Green's function method \cite{Bu:2021clf}. Formally, we have
\begin{align}
V_i^{(m)(n)}(r) &= \int_{\infty_2}^{\infty_1} G_X(r,\xi) j_i^{(m)(n)}(\xi) d\xi  \nonumber \\
& = \frac{X_1(r)}{2i\pi r_h^2} \int_{\infty_2}^r X_2(\xi) j_i^{(m)(n)} (\xi) d\xi + \frac{X_2(r)}{2i\pi r_h^2} \int_r^{\infty_1} X_1(\xi) j_i^{(m)(n)}(\xi) d\xi, \nonumber \\
L_i^{(m)(n)}(r) &= \int_{\infty_2}^{\infty_1} G_X(r,\xi) l_i^{(m)(n)} (\xi) d\xi \nonumber \\
& = \frac{X_1(r)}{2i\pi r_h^2} \int_{\infty_2}^r X_2(\xi) l_i^{(m)(n)} (\xi) d\xi + \frac{X_2(r)}{2i\pi r_h^2} \int_r^{\infty_1} X_1(\xi) l_i^{(m)(n)}(\xi) d\xi,  \label{Vi_Li_mn}
\end{align}
where $X_1(r)$ and $X_2(r)$ are two linearly independent solutions for the homogeneous part of \eqref{dynamical_eom_perturb}. Specifically, $X_1$ and $X_2$ are taken as
\begin{align}
X_1(r) = -\frac{1}{2} \log \frac{r^2-r_h^2}{r^2 +r_h^2} + \i \pi, \qquad \qquad X_2(r) = -\frac{1}{2} \log \frac{r^2-r_h^2}{r^2 +r_h^2}.
\end{align}

\subsection{Computation of the boundary action}\label{fifi}

With the perturbative solutions obtained in section \ref{perturb_solutions}, we are ready to calculate the partially on-shell bulk action \eqref{pos_action}.

Notice that $V_\mu^{(0)}$ and $L_\mu^{(0)}$ are linear in the boundary data $A_{\s\mu}$, $C_{\s\mu}$, while $V_\mu^{(1)}$ and $L_\mu^{(1)}$ are quadratic in the boundary data $A_{\s\mu}$, $C_{\s\mu}$. This fact implies that the linear solutions $V_\mu^{(0)}, L_\mu^{(0)}$ are sufficient in generating quadratic terms, cubic terms and non-anomalous quartic terms in the boundary effective action. Therefore, in order to capture anomalous quartic terms in the boundary action, we have to construct nonlinear solutions $V_\mu^{(1)}$ and $L_\mu^{(1)}$. \\

$\bullet$ \textbf{The quadratic Lagrangian $\mathcal L_{eff}^{(2)}$}

The quadratic Lagrangian is simply obtained by computing
\begin{align}
\mathcal L_{\rm eff}^{(2)} = \int_{\infty_2}^{\infty_1} dr \sqrt{-g} \left( - \frac{1}{4} \mathds{V}_{MN} \mathds{V}^{MN} - \frac{1}{4} \mathds{L}_{MN} \mathds{L}^{MN} \right)\bigg|_{V_\mu \to V_\mu^{(0)}, \,\,L_\mu \to L_\mu^{(0)}} + \mathcal L_{\rm bdy},
\end{align}
where $\mathcal L_{\rm bdy}$ is Lagrangian associated with the boundary action \eqref{S_bdy}, \textit{i.e.}, $S_{\rm bdy} = \int d^4x \mathcal L_{\rm bdy}$. Apart from a slight difference in the boundary term, namely the third term in Eq.~\eqref{S_bdy}, the quadratic Lagrangian was previously computed to second order in derivatives of boundary data \cite{Glorioso:2018mmw,deBoer:2018qqm,Bu:2020jfo}. Recycling those results and truncating them to order $\mathcal O(\varepsilon^2)$ in the MHD regime \eqref{MHD_regime}, we eventually produce the results summarized in Eq.~\eqref{L2}.\\

$\bullet$ \textbf{The cubic Lagrangian $\mathcal L_{eff}^{(3)}$}

For the nonlinear corrections $V_\mu^{(1)}$ and $L_\mu^{(1)}$, the AdS conditions will be imposed in such a way that they both vanish at $r=\infty_\s$. As a result, the nonlinear solutions $V_\mu^{(1)}$ and $L_\mu^{(1)}$ are not needed for calculation of those cubic terms in the boundary action. So, the cubic Lagrangian $\mathcal L_{eff}^{(3)}$ is simply the bulk Chern-Simons term with the linear solutions plugged in
\begin{align}
\mathcal L_{\rm eff}^{(3)} = \int_{\infty_2}^{\infty_1} dr  \sqrt{-g} ~2 \kappa ~\epsilon^{MNPQR} \left[ 3 L_M^{(0)} \nabla_N V_P^{(0)} \nabla_Q V_R^{(0)} + L_M^{(0)} \nabla_N L_P^{(0)} \nabla_Q L_R^{(0)} \right]. \label{L3_radial_integral}
\end{align}

Indeed, it turns out that we only need $V_\mu^{(0)(0)}$ and $L_\mu^{(0)(0)}$ for getting cubic action up to the order $\mathcal O(\varepsilon^2)$ in the MHD regime \eqref{MHD_regime}. With the analytical solutions \eqref{Vmu_Lmu_00}, we are able to analytically work out the radial integral in \eqref{L3_radial_integral},
\begin{align}
\mathcal L_{\rm eff}^{(3)} & = 12 \kappa \epsilon^{ijk}(A_0 C_{\a i} \partial_j A_k + A_0 C_i \partial_j A_{\a k} + A_{\a 0} C_i \partial_j A_k + A_{\a i} C_j \partial_0 A_k) \nonumber \\
& + 4 \kappa \epsilon^{ijk}(C_0 C_i \partial_j C_{\a k} +  C_0 C_{\a i} \partial_j C_k + C_{\a 0} C_i \partial_j C_k + \epsilon^{ijk} C_{\a i} C_j \partial_0 C_k) \nonumber \\
& = 12 \kappa \left(A_0 B_i C_{\a i} + A_0 C_i B_{\a i} + A_{\a 0} C_i B_i + \epsilon^{ijk} A_{\a i} C_j \partial_0 A_k \right) \nonumber \\
& + 4 \kappa (C_0 C_i \mathcal B_{\a i} +  C_0 C_{\a i} \mathcal B_i + C_{\a 0} C_i \mathcal B_i + \epsilon^{ijk} C_{\a i} C_j \partial_0 C_k).
\label{L3_AC}
\end{align}
In order to make comparison with \cite{landry2022systematic}, we rewrite the third line of \eqref{L3_AC} by plugging in the expression $C_{\a i} = \mathcal A_{\a i} + \partial_i \varphi_{\a}$ and $C_i = \mathcal A_i + \partial_i \varphi$. Moreover, integration by parts has also to be used to rewrite the cubic Lagrangian \eqref{L3_AC} in the form of \cite{landry2022systematic}, see \eqref{LL1}. Eventually, the expression \eqref{L3_AC} is recast into the form quoted in \eqref{L3}. Nevertheless, under the KMS transformation \eqref{KMS2}, the invariance of \eqref{L3_AC} is more transparent than that of \eqref{L3}. \\

$\bullet$ \textbf{The non-anomalous quartic Lagrangian $\mathcal L_{\rm normal}^{(4)}$}

Recall that the bulk action \eqref{S0} contains quartic terms for the vector field $V_M$, \textit{i.e.}, the $\alpha_{1,2}$-terms. So, we can generate quartic terms for the boundary action using the linear solutions $V_{\mu}^{(0)}$
\begin{align}
\mathcal L_{\rm normal} = \int_{\infty_2}^{\infty_1} dr \sqrt{-g} \left[ \alpha_1 \left( \mathds{V}_{MN} \mathds{V}^{MN} \right)^2 + \alpha_2 \mathds{V}_{AB} \mathds{V}^{BC} \mathds{V}_{CD}\mathds{V}^{DA} \right]\bigg|_{V_\mu \to V_\mu^{(0)}} .\label{L4_normal_integral}
\end{align}
While \eqref{L4_normal_integral} seems to contain plenty of terms, truncating it at the order $\mathcal O(\varepsilon^2)$ in the MHD regime \eqref{MHD_regime} does significantly reduce this number. Eventually, evaluating the radial integral \eqref{L4_normal_integral} gives the non-anomalous quartic Lagrangian quoted in \eqref{L4_normal}. Here, we stress that these quartic terms are not directly related to the chiral anomaly.\\

$\bullet$ \textbf{The anomalous quartic Lagrangian $\mathcal L_{\rm anom}^{(4)}$}

The last part of the boundary action we derived corresponds to the quartic terms fully generated by the chiral anomaly. Through the Chern-Simons term in \eqref{S0}, we have
\begin{align}
\mathcal L_{\rm anom}^{(4)} = \int_{\infty_2}^{\infty_1} dr \sqrt{-g} ~\frac{\kappa}{2} \epsilon^{MNPQR} & \left[ 3 L_M^{(1)} \mathds V_{NP}^{(0)} \mathds V_{QR}^{(0)} + 6 L_M^{(0)} \mathds V_{NP}^{(1)} \mathds V_{QR}^{(0)} + L_M^{(1)} \mathds L_{NP}^{(0)} \mathds L_{QR}^{(0)} \right. \nonumber \\
&\, \left. + 2 L_M^{(0)} \mathds L_{NP}^{(1)} \mathds L_{QR}^{(0)} \right] \nonumber \\
= \int_{\infty_2}^{\infty_1} dr \sqrt{-g}~ 3 \kappa \epsilon^{MNPQR} & \left[ 2V_M^{(1)} \nabla_N V_P^{(0)} \nabla_Q L_R^{(0)} + L_M^{(1)} \nabla_N V_{P}^{(0)} \nabla_Q V_R^{(0)} \right. \nonumber \\
& \, \left. + L_M^{(1)} \nabla_N L_P^{(0)} \nabla_Q L_R^{(0)} \right] ,\label{L4_anom_integral}
\end{align}
where in order to get the second equality we have integrated by parts. Inevitably, the computation of $\mathcal L_{\rm anom}^{(4)}$ requires to solve the nonlinear corrections $V_\mu^{(1)}$ and $L_\mu^{(1)}$.

With the nonlinear solutions presented in \eqref{V0_L0_mn} and \eqref{Vi_Li_mn}, the evaluation of \eqref{L4_anom_integral} is straightforward although a bit tedious, since it involves multiple radial integrals. Skipping the details, we present the final result
in a more compact form
\begin{align}
\mathcal L_{\rm anom}^{(4)} =& M_{abcd} \left\{  \theta_1 \left[ (\vec \nabla \times \vec \Psi^{a})_i (\vec \nabla \times \vec \Psi^{b})_i \Psi_0^{c} \Psi_{{\rm a} 0}^{d} + (\vec \nabla \times \vec \Psi^{a})_i (\vec \nabla \times \vec \Psi_{{\rm a}}^{b})_i \Psi_{0}^{c} \Psi_{0}^{d} \right]\right. \nonumber \\
& \left.- \theta_2 \epsilon_{ijk} (\vec \nabla \times \vec \Psi^{a})_k \partial_{0} \Psi_{i}^{b} \Psi_{0}^{c} \Psi_{{\rm a} j}^{d} \right. \nonumber \\
& \left. + \i \theta_3 \left[ (\vec \nabla \times \vec \Psi^{a})_i ( \vec \nabla \times \vec \Psi^{b})_i \Psi_{{\rm a} k}^{c} \Psi_{{\rm a} k}^{d} - (\vec \nabla \times \vec \Psi^{a})_j (\vec \nabla \times \vec \Psi^{b})_k \Psi_{{\rm a}  j}^{c} \Psi_{{\rm a} k}^{d} \right] \right. \nonumber \\
& \left. - \theta_3 \beta \left[ (\vec \nabla \times \vec \Psi^{a})_i (\vec \nabla \times \vec \Psi^{b})_i \partial_{0} \Psi_{k}^{c} \Psi_{{\rm a} k}^{d} - (\vec \nabla \times \vec \Psi^{a})_j (\vec \nabla \times \vec \Psi^{b})_k \partial_{0} \Psi_{j}^{c} \Psi_{{\rm a} k}^{d} \right] \right\} \label{L4_anom_MPsi}
\end{align}
where values of $\theta_{1,2,3}$ are presented in \eqref{theta_values}. The doublet $\Psi^a_\mu$ is used to represent $(A_\mu, C_\mu)$ while the doublet $\Psi^a_{\a\mu}$ denotes $(A_{\a\mu}, C_{\a\mu})$. More precisely, we have
\begin{align}
\Psi^1_\mu = A_\mu, \qquad \Psi^2_\mu = C_\mu, \qquad \Psi^1_{\a\mu} = A_{\a\mu}, \qquad \Psi^2_{\a\mu} = C_{\a\mu}.
\end{align}
We remind that the vector notations of the form $\vec \Psi^a$ collectively refer to the spatial components of $\Psi^a_\mu$. In addition, the tensor $M_{adcd}$ is defined as
\begin{align}
M_{abcd} = \delta_{ab} \delta_{cd} + \delta_{ac} \delta_{bd} + \delta_{ad} \delta_{bc} - 2 \delta_{abcd}.
\end{align}
Interestingly, the anomalous quartic Lagrangian \eqref{L4_anom_MPsi} is symmetric under the exchange $A \leftrightarrow C$. More explicitly, we rewrite \eqref{L4_anom_MPsi} to the one quoted in \eqref{L4_anom1}.

\section{Summary and Outlook}
In this work, we have constructed a holographic model for chiral anomalous magnetohydrodynamics in $3$+$1$ dimensions. This was achieved by combining the introduction of a 5D bulk Chern-Simons term with modified boundary conditions for the EM U(1) bulk gauge field. By using holographic Schwinger-Keldysh techniques, the SK effective action of the boundary field theory is derived and proved to exactly match the field theory construction proposed by Landry and Liu \cite{landry2022systematic}. Importantly, we generalize the SK effective action in Landry and Liu \cite{landry2022systematic} where the background axial gauge field was set to zero for convenience. Here, we provide a complete action also in presence of the latter.

Apart from confirming the validity of the EFT action in \cite{landry2022systematic} and the correct nature of the holographic model, our results open the path towards the exploration of CAMHD in the strong coupling and strong magnetic field regime using holography. In particular, it would be interesting (I) to confirm the existence of a \textit{chiral magnetic electric separation wave} and study its dynamics in the strong $B$ limit and (II) to explore the chiral instabilities expected at nonlinear level. A first step would be to obtain the quasinormal mode spectrum of the system and study the real-time dynamics in this holographic model along the lines of similar studies in holographic chiral systems \cite{Grieninger:2023myf,Ghosh:2021naw}.

At the order in the magnetic field considered in this work, we can already anticipate some results regarding the \textit{chiral magnetic electric separation wave}.\footnote{We thank Xin-Meng Wu for this suggestion.} In particular, as derived in \cite{landry2022systematic} (see eqn (6.3) therein), its dispersion relation reads
\begin{align}
\omega = - \i\, \Gamma + v_z k_z + \mathcal O(k_{z,\perp}^2), \qquad \Gamma = \frac{(2c B)^2 r_\parallel}{\chi_5}, \qquad v_z = \frac{4c B r_{\parallel} \lambda_{\parallel}}{\chi_5}.
\end{align}
Recall that $c$ appears in \eqref{LL1} and is related to the anomaly coefficient $\kappa$ appearing in \eqref{S0} by $c = - 6 \kappa$.
Using our holographic computation, we can derive closed-form expressions for the various parameters entering therein. In particular, we obtain
\begin{align}
		&\chi_5 \equiv  \frac{\partial n_5}{\partial \mu_5} = \frac{2 \pi^2}{\beta^2} + \theta_1 B^2 + \theta_1 \mathcal B^2, \nonumber \\
		&(r^{-1})_\parallel =  \frac{ \pi}{\beta}  +  \frac{8 \beta }{\pi}(2 \alpha_{1} +  \alpha_2)\left(4 A_0^2 - \frac{ \beta^2 B^2}{ \pi^2}\right)   + \frac{8 \beta^3}{\pi^3} \alpha_{2} B^2 + 2 \theta_3 \beta B^2 +  \theta_3 \beta(\mathcal B^2 + \frac{\mathcal B_i \mathcal B_j}{B_i B_j} B^2 ), \nonumber \\
		&\lambda_\parallel =  4 \theta_3 \beta B_i \mathcal B_i.
\end{align}
In absence of background axial gauge field, $\mathcal{B}_i=0$, we have $\lambda_\parallel=0$ and, hence, $v_z=0$. On the other hand, we find that
\begin{equation}
    \Gamma= \frac{2 c^2}{\pi^3 T^3} B^2 +\dots
\end{equation}
in the small magnetic field limit and in the limit of zero vector chemical potential ($A_0=0$). Therefore, we can already conclude that, in absence of background axial field, and in the limit of small magnetic field, the \textit{chiral magnetic electric separation wave} is not a propagating wave but rather a purely decaying damped excitation. It would be interesting to perform a full quasinormal mode analysis in the large magnetic field to ascertain whether this changes in such a limit.

We expect that the introduction of a background axial gauge field will have several physical consequences that can be systematically studied using our framework. Despite a complete study on this point is beyond the scope of this work, we can anticipate one interesting effect caused by the presence of a background axial gauge field and related to the dynamics of the chiral magnetic electric separation wave discussed above. In particular, as evident from the equations above, the background axial magnetic field $\mathcal B_i$ significantly affects the wave's dispersion relation, modifying both the damping rate and the wave velocity. In fact, in the regime of weak magnetic field, the wave velocity is exactly zero if the background axial gauge field is turned off. We do expect more physical consequences of this sort.\color{black}

We leave this and other interesting questions for the future.

\appendix

\section{Symbols and notations}\label{note}

\begin{table}[h!]
\begin{tabular}{||c c|| ||c c||}
 \hline
 Symbol & Physical meaning & Symbol & Physical meaning  \\ [0.5ex] 
 \hline\hline
 $T = 1/\beta$ & temperature & $e$ & electric charge  \\ 
 \hline
 $\psi$ & Dirac field & $\mu$ & magnetic permittivity  \\ 
 \hline
 $a_\mu$ & e.m. gauge potential & $a_{\a\mu}$ & noisy counterpart of $a_\mu$  \\ 
 \hline
 $\mathcal A_\mu$ & axial gauge potential & $\mathcal A_{\a\mu}$ & noisy counterpart of $\mathcal A_\mu$  \\ 
 \hline
 $\phi$ & vector diffusive field & $\phi_\a$ & noisy counterpart of $\phi$  \\ 
 \hline
 $\varphi$ & axial diffusive field& $\varphi_\a$ & noisy counterpart of $\varphi$  \\ 
 \hline
 $A_\mu$ & combination $a_\mu + \partial_\mu \phi$ & $A_{\a\mu}$ & noisy counterpart of $A_{\a\mu}$  \\ 
 \hline
 $C_\mu$ & combination $\mathcal A_\mu + \partial_\mu \varphi$ & $C_{\a\mu}$ & noisy counterpart of $C_\mu$ \\ 
 \hline
 $A_0$ & vector chemical potential & $A_{\a0} $ & noisy counterpart of $A_0$ \\
 \hline
 $C_0 \equiv \mu_5$ & axial chemical potential & $C_{\a0}$ & noisy counterpart of $C_0$ \\
 \hline
 $\vec B$ & vector magnetic field & $\vec B_\a$ & noisy counterpart of $\vec B$ \\
 \hline
 $\vec{ \mathcal{B} }$ & axial magnetic field & $\vec{ \mathcal{B} }_\a$ & noisy counterpart of $\vec{ \mathcal{B} }$ \\
 \hline
 $\vec E$ & vector electric field & $\vec E_\a$ & noisy counterpart of $\vec E$ \\
 \hline
 $\vec{\mathcal{E}}$ & axial electric field & $\vec{\mathcal{E}}_\a$ & noisy counterpart of $\vec{\mathcal{E}}$ \\
 \hline
 $J^\mu = (J^0 \equiv n,\, J^i)$ & vector 4-current& $J_5^\mu = (J_5^0 \equiv n_5,\, J_5^i)$ & axial 4-current \\
 \hline
 $V_M$ & bulk vector field & $L_M$ & bulk axial field \\
 \hline
 $\mathfrak J_{\s\mu}$ & normalizable modes of $V_M$ & $\mathfrak L_{\s\mu}$ & normalizable modes of $L_M$ \\
 \hline
 $\mathds V_{MN}$ & field strength of $V_M$ & $\mathds L_{MN}$ & field strength of $L_M$ \\
 \hline
 $F^{\mu\nu}$ & electromagnetic tensor & $\mathcal{J}^\mu$ & the combination $\partial_\nu F^{\nu\mu} + J^\mu$  \\
 \hline
\end{tabular}
\caption{Symbols and notations used in this paper.}
\label{table1}
\end{table}

\section*{Acknowledgments}
We thank Michael Landry, Hong Liu, Sebastian Grieninger, Alberto Salvio, Stefano Lionetti, Danny Brattan and Xin-Meng Wu for useful discussions.
MB acknowledges the support of the Shanghai Municipal Science and Technology Major Project (Grant No.2019SHZDZX01) and the sponsorship from the Yangyang Development Fund. YB and XS were supported by the National Natural Science Foundation of China (NSFC) under the grant No. 12375044.

\bibliographystyle{JHEP}
\bibliography{reference}
\end{document}